\documentclass[12pt]{iopart}

\usepackage{iopams}  
\usepackage{graphicx}
\usepackage{dcolumn}
\usepackage{bm}
\usepackage{soul,color}
\usepackage[dvipsnames]{xcolor}
\usepackage{hyperref}
\usepackage{amsmath}
\usepackage{tabularx}

\begin{document}

\title[]{Analyzing coherent phonon mode-conversion in gradient superlattices with atomistic wave-packet simulations}

\author{
    Evan Wallace Doe$^1$\footnote{Corresponding author: edoe@unr.edu},
    Theodore Maranets$^1$,
    Yan Wang$^1$\footnote{Corresponding author: yanwang@unr.edu}
}
\address{$^1$Department of Mechanical Engineering, University of Nevada, Reno, Reno, NV, 89557, USA }

\begin{abstract}

In this study, we have used atomistic phonon wave-packet simulations to investigate the manifestation of coherent phonons and phonon transmission in gradient superlattices (SL) based on ordered arrangements of varied SL period sizes. We specifically explore how coherent mode-conversion in these quasi-periodic structures changes as function of three key structural parameters: (1) the number of distinct period sizes, (2) the number of periods present for each distinct period size, and (3) the arrangement of period sizes in either an ascending or descending arrangement. Comparisons to periodic SLs and aperiodic SLs are highlighted, revealing that coherent phonons in gradient SLs generally exhibit behaviors characteristic of intermediate states between the fully ordered and disordered structures. Interestingly, changes to the short-range order of GMLs does not significantly influence transmission, indicating that long-range disorder is far more important to coherent mode-conversion. Our results indicate that manipulating the long-range disorder of interfaces could be an effective strategy to tailor phonon thermal conductivity of SL architectures.

\end{abstract}

\maketitle

\section{Introduction\label{sec:introduction}}

Artificial materials possessing secondary periodicity can substantially disrupt the phonon wave functions at nano-scale dimensions, leading to unique thermophysical properties \cite{maldovan2015phonon, volz2016nanophononics, xie2018phonon, zhang2021coherent, anufriev2021review}. Superlattices (SLs) are one such structure that can stimulate new wave-like phonon modes from scattering resonances at periodically arranged interfaces \cite{latour2014microscopic, latour2017distinguishing, maranets2024influence, maranets2024prominent, maranets2025how, maranets2025role}. Conventional modeling methodologies like the Boltzmann transport equation that characterize phonon transport as particle-like overlook phonon wave phenomena like interference, coherence, and localization which emerge in SLs and similar structures \cite{simkin2000minimum, ravichandran2014crossover, luckyanova2012coherent, wang2014decomposition, anufriev2016reduction, puurtinen2016low, cui2024spectral}. As a result, current techniques aimed at understanding thermal transport in these materials are restricted by their inadequate consideration of phonon coherence. 

The manifestation of wave-like phonon behaviors, termed ``phonon coherence", offer a pathway to design materials with tailored thermal properties via phonon wave manipulation. Phonon coherence is dictated by the comparison of the phonon spatial coherence length to the characteristic length of a material \cite{latour2014microscopic,latour2017distinguishing,maranets2024influence}. When the coherence length exceeds the spacing between interfaces in SLs, wave-like coherent phonons that propagate with high transmission predominate. For coherence lengths shorter than the interface spacings, the interface scattering of particle-like incoherent phonons is the dominant transport mechanism. Thermal conductivities of SL architectures can be engineered through structural variations, such as aperiodic layering, which introduce disorder into the periodicity and have been shown to disrupt phonon coherence and stimulate localization \cite{maranets2024influence, maranets2024prominent, maranets2025how, maranets2025role, wang2015optimization, juntunen2019anderson, chowdhury2020machine, ma2020dimensionality, hu2021direct, maranets2025phonon, guo2021thermal}. Chowdhury $et$ $al.$ \cite{chowdhury2020machine} leveraged machine learning to demonstrate that the thermal conductivity of aperiodic SLs attains a minimum at a moderate level of randomness. These findings demonstrate that randomization of SL periodicity is limited and that truly tailoring SLs must rely on alternative methods. One possible avenue that has been minimally investigated in the literature is SLs possessing quasi-periodic layer arrangements.

The architecture of periodic SLs are inherently orderly, consisting of uniform layers of two alternating materials, whereas aperiodic SLs or random multi-layers (RMLs) are inherently disordered, with random layer thicknesses. In this study, we examine quasi-periodic SLs, in which the structure progressively increases or decreases the layer lengths from one end to the other. These gradient multilayered (GML) structures are orderly, but possess deterministic disorder, which distinguishes them from other architectures. GMLs bridge the gap between periodic and aperiodic SLs as these systems contain short-range order but long-range disorder. It has been observed that simultaneous structural order and disorder in crystalline solids significantly influences phonon dynamics \cite{guo2021thermal, zhang2025anomalous, zhang2026decoding, xie2022impacts}. For GMLs, Guo $et$ $al.$ demonstrated through non-equilibrium Green's function that the deterministic disorder in a Si/Ge GMLs achieves a minimum thermal conductivity due to partial phonon localization \cite{guo2021thermal}. While this represents a significant advance in understanding how varying levels of disorder affect phonon coherence, it does not attain the deeper insights into the wave behaviors that atomistic phonon wave‑packet simulations can reveal.

An accurate picture of phonon coherence based on the wave nature of phonons is challenging, to say the least. Phonon coherence within SLs and other metamaterials is significantly influenced by the spatial coherence length of phonons \cite{maranets2024influence}. The investigation into how structural modifications and changes in spatial coherence length affect phonon coherence has thus far been confined to theoretical analysis, as current experimental techniques lack the precision required for accurate characterization of phonon coherence \cite{luo2013nanoscale, minnich2015advances, lindsay2019perspective}. This dependence on the spatial coherence length is the primary characteristic that distinguishes the wave‑packet approach from alternative methodologies. The robust wave-packet methodology offers precise information for the investigation of phonon coherence as a function of spatial coherence length, phonon mode, and wave-vector. \cite{latour2017distinguishing, maranets2024influence, maranets2024prominent, maranets2025how, latour2017distinguishing, maranets2025phonon, schelling2002phonon, tian2010phonon, liang2017phonon, maranets2023ballistic}. With these essential parameters of phonon waves described, wave-packet simulations can provide unique comprehensive understanding of coherent phonon transport in engineered SLs \cite{latour2014microscopic, latour2017distinguishing, maranets2024influence, maranets2024prominent, maranets2025how, maranets2025role, maranets2025phonon}. We have previously used wave-packet simulations to validate existing hypotheses and produce novel insights into phonon coherence in periodic and aperiodic SLs \cite{maranets2024influence,maranets2024prominent,maranets2025how,maranets2025role}. Now, we seek to connect our prior investigations to a new wave-packet study on gradient SLs or GMLs that are quasi-periodic.

\section{Methodology \label{sec:methodology}}

\subsection{Material system\label{sec:material_system}}

We use the same model material system as our previous studies on periodic and aperiodic SLs to facilitate direct comparison with the results for the gradient SL. The material system utilized is based on a Lennard-Jones solid Argon structure. The two layer materials composing the SLs have molecular weights of 40 g mol$^{-1}$ and 90 g mol$^{-1}$ and are referred to as m40 and m90, respectively. Both materials have identical force constants, resulting in zero elastic contrast. To mimic covalently bonded semiconductor SLs like Si/Ge and GaAs/AlAs that have been studied experimentally, we utilize a well depth of $\varepsilon = 0.1664$ eV which is 16 times the original value for solid Argon \cite{wang2014decomposition, wang2015optimization, chakraborty2017ultralow, landry2008complex, chakraborty2020complex, maranets2023lattice}. The zero-crossing distance is $\sigma = 3.4$ \AA  and the cutoff radius is set to $2.5\sigma$. The lattice constant corresponding to 1 UC for this material system is 5.27 \AA.

\subsection{Superlattice architectures \label{sec:SL_architectures}}

\begin{figure}
    \centering
    \includegraphics[width=0.9\textwidth]{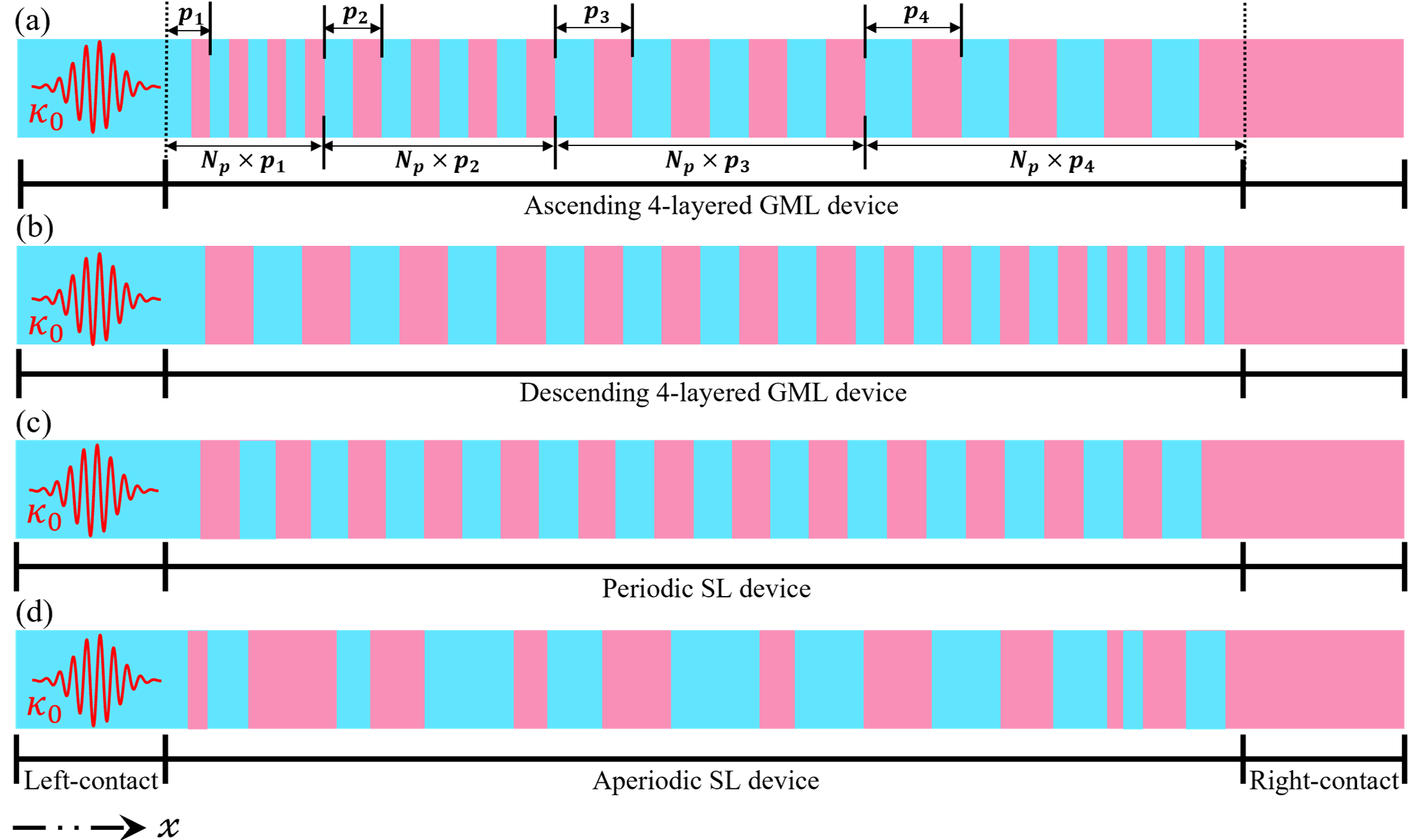}
    \caption{Schematic illustrations of wave-packet simulations domains for: (a) an ascending gradient multilayer (GML) device with 4 different SL period sizes and 4 periods for each period size, (b) a descending GML device with 4 different SL period sizes and 4 periods for each period size, (c) a periodic SL device, and (d) an aperiodic SL device. $k_0$ denotes the central wavevector of the incident LA-mode incoherent phonon wave-packet. The phonon wave-packet is generated in the left contact and is allowed to propagate into the device over the duration of the simulation. The total energies of the left and right contacts and the device are monitored to compute the energy transmission across the device using Eqn.~\ref{eqn:transmission_calc}. Illustrations of the wavelength and spatial coherence length for the wave-packet are not to scale.}
    \label{fig:SL-GML-RML}
\end{figure}

We investigate SL architectures of periodic, gradient, and aperiodic layer arrangements. Periodic SLs are characterized by a repeated pattern of alternating m40 and m90 layers of uniform thickness. Gradient SLs, to be referred to as gradient multilayers (GML) in this study, possess layers of varying thickness, but in an ordered arrangement. Lastly, aperiodic SLs or random multilayers (RML) contain layers of varying thickness in a disordered arrangement.

Three structural parameters characterize GML configurations: (1) the number of different period sizes $N_{s}$, (2) the number of periods for each period size $N_p$, and (3) the ordering of the period sizes in either an ascending or descending pattern. We represent these parameters in the following equation for GML device length $L$.
\begin{equation}
L = \sum_{i=1}^{N_s} d_i \cdot N_p
\label{eqn:gml_equation}
\end{equation}
where $d_{i}$ signifies SL period sizes. $d_{i+1}>d_{i}$ corresponds to a GML device of ascending layer thickness while $d_{i+1}<d_{i}$ corresponds to a GML device of descending layer thickness. We examined GML configurations with variation in the three structural parameters to assess all factors that could affect phonon transmission and coherent mode-conversion in GMLs. Schematic illustrations of these parameters and comparative periodic SL and RML devices in our wave-packet simulation domains are presented in Fig.~\ref{fig:SL-GML-RML}. For the ascending GML, the initial device period comprised 2 UCs of m40/m90; upon reaching the specified $N_p$, the system transitions to the next period size. Each GML structure effectively consists of multiple periodic SLs with the same $N_p$ before an increase or decrease in the periodicity occurs. Fig.~\ref{fig:SL-GML-RML}a depicts an ascending GML with $N_s=4$, showing that after the conclusion of the fourth period, the layer thickness increases incrementally by 1 UC. This process is repeated for a selected set of period sizes as determined by $N_s$ before the device length is terminated. Notably, the number of distinct period sizes equals the number of distinct gradient layers for each period size. We examined $N_s$ values of 3, 4, 5, 6, and 7 for $N_p=4$, $N_p=8$, and $N_p=16$. Our RML devices were constructed by disassembling GML devices and reassembling them in a random sequence. This methodology ensures that the structures retain the same number of material layers with identical sizes as the GML structure. This results in states that are more ordered than the configurations in our previous studies, but importantly more disordered than the GML structures.

\subsection{Wave-packet simulations\label{sec:wave-packet simulation}}

We performed atomistic phonon wave-packet simulations based on the method developed by Schelling $et$ $al.$ \cite{schelling2002phonon} and the techniques of our prior studies \cite{maranets2024influence,maranets2024prominent,maranets2025how,maranets2025role}. Specifically, we probe how longitudinal-acoustic (LA) incoherent phonons behave when propagating through periodic SL, GML, and RML devices.

To initiate a phonon wave-packet simulation, a configuration of atomic displacements and velocities with the literal form of a wave-packet is generated within an equilibrium lattice at zero-temperature. The initial displacement and velocity of the $i$th atom within the $n$th UC are specified by the following equations:
\begin{equation}
u_{i,n} = \frac{A_{i}}{\sqrt{m_{i}}}\varepsilon_{k_{0},i}\exp{(i[k_{0}\cdot(x_{n} - x_{0})-\omega_{0}t]})\exp{(-4(x_{n}-x_{0}-v_{g0}t)^{2}/l_{c}^{2})},\qquad 
\label{eqn:pwp_disp}
\end{equation}
\begin{equation}
v_{i,n} = \frac{\partial u_{i,n}}{\partial t},\qquad
\label{eqn:pwp_vel}
\end{equation} 
where $m_{i}$ represents the mass of atom $i$, and $x_{n}$ signifies the position of the $n$th unit cell along the axis of wave-packet propagation. The eigenvector, frequency, and group velocity associated with the wavevector $k_{0}$ are denoted by $\varepsilon_{k_{0},i}$, $\omega_{0}$, and $v_{g0}$. Since the wave-packet is generated in the left-contact composed of m40, these phonon properties are derived from the bulk dispersion relation of m40. User-specified parameters include the wave amplitude $A_{i}$, the initial position $x_{0}$ from which the wave packet originates, and the spatial coherence length $l_{c}$. The real parts of Eqns.~\ref{eqn:pwp_disp} and \ref{eqn:pwp_vel} evaluated at time $t=0$ are used as the perturbation values for wave-packet initialization. Following the stimulation of a phonon wave-packet, the simulations are ran in a micro-canonical ensemble with periodic boundary conditions implemented along all three spatial dimensions. We have utilized the LAMMPS software \cite{thompson2022lammps} for these simulations. 

The computation of the wave-packet transmission across SL devices is performed using the following equation:
\begin{equation}
    \mathcal{T}=\frac{E_{rc,final}}{E_{lc,initial}},\qquad
    \label{eqn:transmission_calc}
\end{equation} 
where the final and initial atomic energies of the right (rc) and left (rc) contacts are denoted by $E_{rc,final}$ and $E_{lc,initial}$, respectively. The parameters of the simulations are set so that all incident energies have been transmitted or reflected through by the end of the simulation duration. 

The device configurations implemented in this study have lengths sufficiently large such that incoherent phonons are sufficiently attenuated by the multitude of interfaces and don't contribute to the overall transmission \cite{maranets2025how}. Thus any observed transmission is a consequence of incoherent phonons mode-converting to coherent phonons. We accentuate opportunities for this effect to occur by implementing a spatial coherence length $l_c$ four times that of the device length $L$ \cite{maranets2024influence}. We note that spatial coherence length $l_c$ varies with temperature \cite{latour2014microscopic} and that temperature-dependence of phonon dynamics in zero-temperature wave-packet simulations could be investigated by changing $l_c$ \cite{maranets2024influence}. However, such effects of temperature are not the focus of this study, but rather how coherent mode-conversion manifests and influences phonon transmission in GMLs.

\section{Results \label{sec:Results}}

\subsection{Number of periods for each period size \label{sec:Number_of_periods}} 

\begin{figure}%
    \centering%
    \includegraphics[width=1\textwidth]{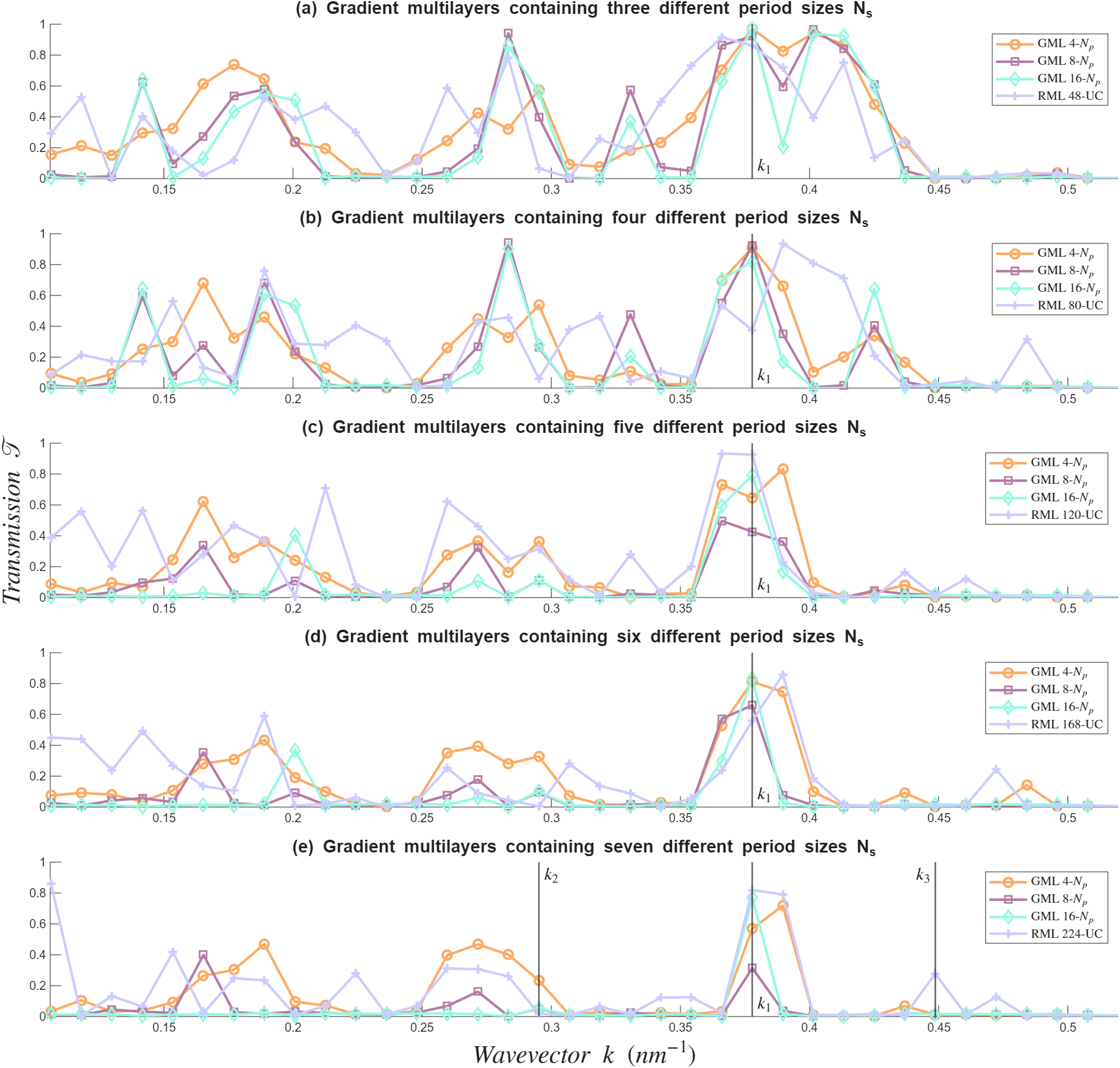}
    \caption{Transmission $\mathcal{T}$ versus wavevector $k$ for longitudinal-acoustic (LA) incoherent phonon wave-packets propagating through gradient multilayer (GML) and random multilayer (RML) devices. Transmission is computed using Eq.~\ref{eqn:transmission_calc}. The GMLs are categorized by the number of distinct period sizes $N_s$ and the number of periods included for each period size $N_p$. For each 4-$N_p$ configuration, we compare the transmission spectra to that of an RML with equivalent length. Select wavevectors $k_1$, $k_2$, and $k_3$ investigated in reciprocal-space wavelet transforms in Fig.~\ref{fig:gml_wavelet} are marked.}
    \label{fig:GML_period_sizes}
\end{figure}

We begin with assessing how the number of periods included in the structure for each distinct period size $N_p$ influences phonon transmission in GMLs. In Fig.~\ref{fig:GML_period_sizes}, we have plotted the transmission spectra of varying $N_p$ for each configuration of the number of distinct period sizes $N_s$ we simulated. Across all structures, low 4-$N_p$ configurations exhibit broad transmission spectra while higher 16-$N_p$ configurations demonstrate more narrow spectra with isolated transmission peaks akin to that observed in the RML. These RML transmission spectra are consistent with the results of our prior research \cite{maranets2024prominent,maranets2025how,maranets2025role}. As $N_s$ is increased, the 4-$N_p$ spectra shifts to a more narrow and peaked spectrum like the 16-$N_p$ though the differences in the overall transmission between different $N_p$ configurations becomes more pronounced. 

\begin{figure}
    \centering
    \includegraphics[width=0.9\textwidth]{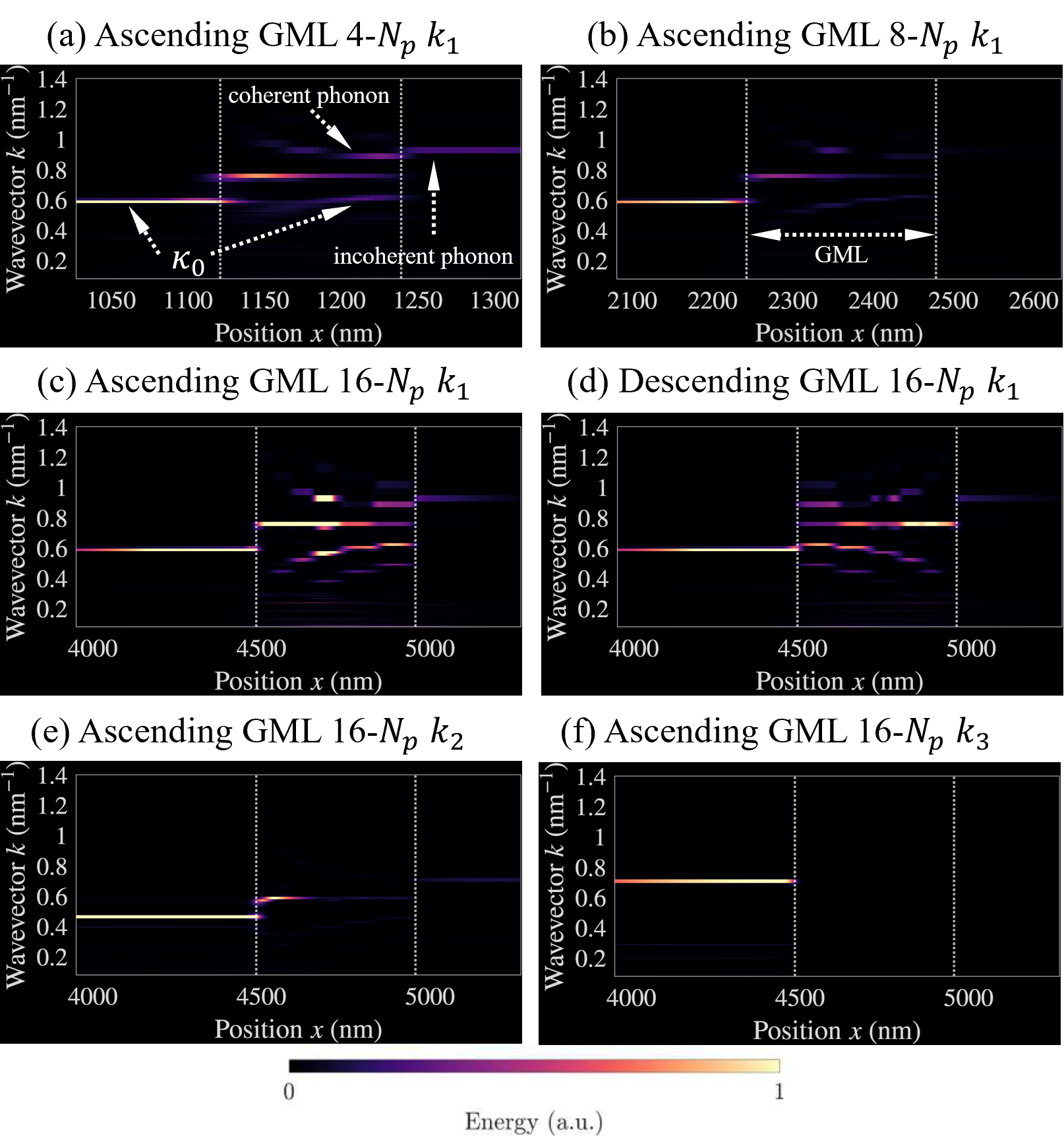}
    \caption{Snapshots of the reciprocal-space wavelet transform for $k_1$, $k_2$, and $k_3$ as the incoherent phonon wave-packets propagate through the GML devices of several configurations of $N_p$ for $N_s=7$. The dotted vertical lines indicate the position and boundaries of the GML device. Arrows and text labels identify the wavevector of the incident incoherent phonon as well as the coherent mode-conversion inside the device \cite{maranets2024prominent}.}
    \label{fig:gml_wavelet}
\end{figure}

To elucidate the mechanisms driving the changes in transmission with varying $N_p$, we performed reciprocal-space wavelet transforms based on the technique by Baker $et$ $al.$ \cite{baker2012application}. The wavelet transform allows us to examine how the phonon wave dynamics evolve in both real-space and reciprocal-space simultaneously, thereby enabling quantification of coherent mode-conversion, the process in which an incoherent phonon converts to a coherent phonon described by the intrinsic phonon spectrum of the SL device. In Fig.~\ref{fig:gml_wavelet}, we provide a comparison of the wavelet transforms for 4-$N_p$, 8-$N_p$, and 16-$N_p$ GML configurations of $N_s$ = 7 at select wavevectors marked in Fig.~\ref{fig:GML_period_sizes}. We first observe that like our previous studies of periodic and aperiodic SLs, prominent transmission through the GML is dictated by the extent of coherent mode-conversion. We also find that this coherent mode-conversion in GMLs varies with $N_p$, explaining the differences in transmission assessed earlier in this section. Specifically, the 4-$N_p$ structures show coherent phonons of broadened wavevector spanning the length of the GML device while coherent phonons in 16-$N_p$ structures have a more narrow wavevector range which changes in a step-wise fashion across the length of the device. These  characteristics observed for the 16-$N_p$ GML configurations align with our prior analyses of the length-dependence of coherent mode-conversion in periodic SLs \cite{maranets2025how}. Mainly, we found that 16 periods is the length at which coherent mode-conversion and Bragg reflection is fully developed in periodic SLs. Considering the GML is effectively a series arrangement of periodic SLs of varying period size, the wavelet transform results for the 16-$N_p$ configurations shown in Fig.~\ref{fig:gml_wavelet} are expected. At 16-$N_p$, the periodic SL domains within the GML device are sufficiently long enough for the incident incoherent phonons to mode-convert to coherent phonons of the periodic SL spectra. Both the narrow wavevectors and the evolution of wavevector in a step-wise fashion aligning with the step-wise changes in period size support this argument. In contrast, the 4-$N_p$ GML configurations are not able to facilitate such effect and so the coherent mode-conversion observed correspond to the intrinsic spectrum of the GML. Comparable characteristics to RML coherent phonons studied in our prior works \cite{maranets2024prominent,maranets2025how} support this assessment. 

\subsection{Number of different period sizes\label{sec:number_of_period_sizes}}

In this section, we analyze the impact of the number of different period sizes $N_s$ on phonon transmission and coherent mode-conversion in GMLs. Figs.~\ref{fig:GML_period_sizes}a-\ref{fig:GML_period_sizes}e show that increases in $N_s$ effectuate (1) an overall reduction in transmission and (2) a transition to a more peaked spectrum comparable to the RML for all $N_p$ configurations.  We also see that strong similarity in the transmission spectra between 4-$N_p$ GMLs and RMLs of equivalent lengths does not change with increases in $N_s$.

\begin{figure}
    \centering
    \includegraphics[width=0.9\textwidth]{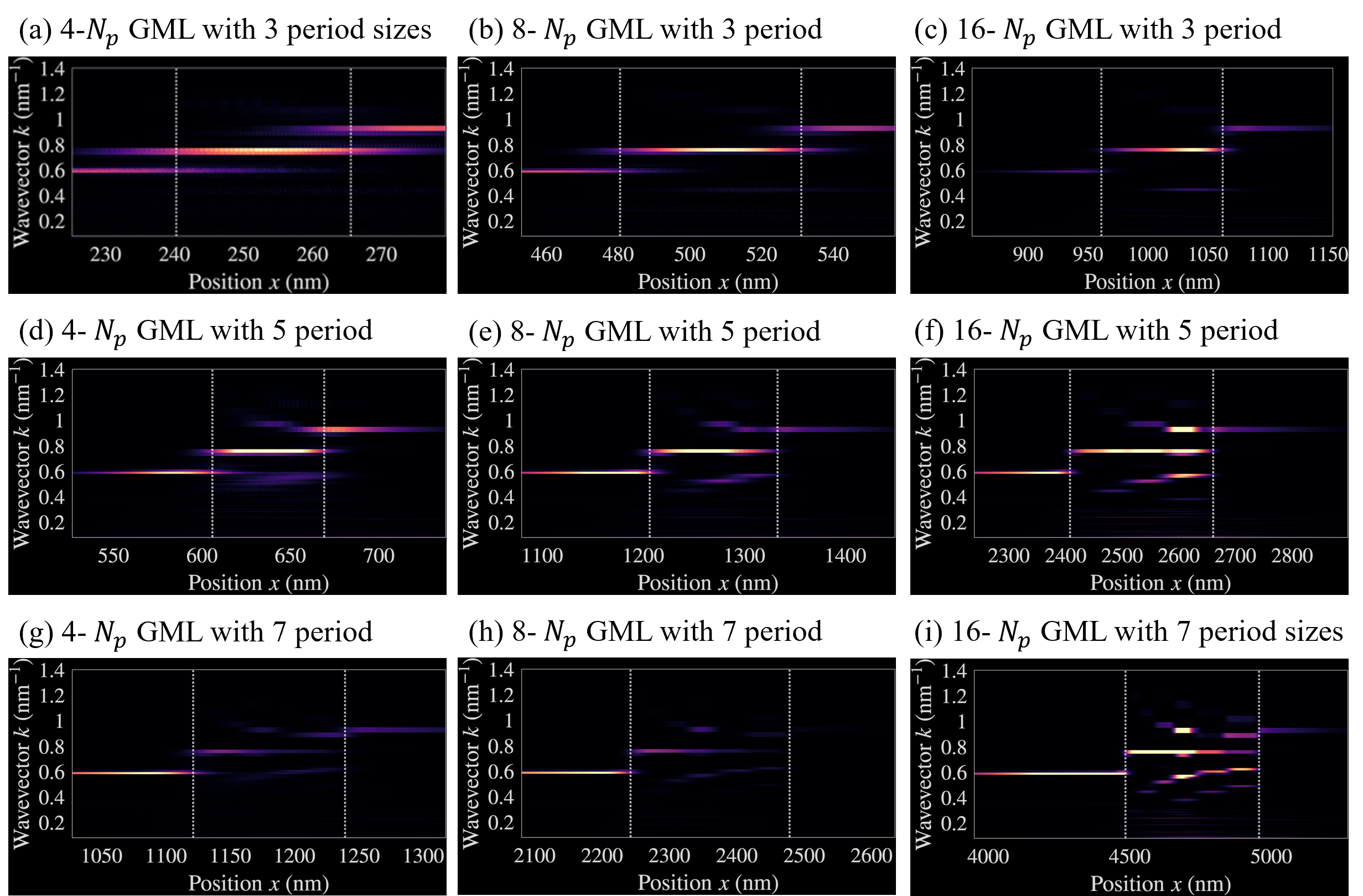}
    \caption{Snapshots of the reciprocal-space wavelet transform for $k_1$ as the incoherent phonon wave-packets propagate through the GML devices with $N_s=3$, 5, and 7 for differing $N_p$. The dotted vertical lines indicate the position and boundaries of the GML device. Arrows and text labels identify the wavevector of the incident incoherent phonon as well as the coherent mode-conversion inside the device \cite{maranets2024prominent}.}
    \label{fig:gml_3-7}
\end{figure}

To understand these effects, we return to analysis of the reciprocal-space wavelet transform data. In Fig.~\ref{fig:gml_3-7}, we present wavelet results for $N_s$ values of 3, 5, and 7 for varying $N_p$. We first observe that weaker transmission for large $N_s$ is associated with reduced coherent mode-conversion. As mode-conversion is attenuated, more phonons across the spectrum exhibit zero transmission, thereby yielding transmission spectra of narrow isolated peaks corresponding to the remaining high transmission modes that are able to mode-convert. Furthermore, we find the features and mode shapes of coherent phonons inside the GML device to change as $N_s$ increases. For $N_s=3$, we see fairly narrow wavevector signatures that are quite comparable to those observed in periodic SLs. However, at $N_s=7$, the observed coherent phonons with broadened wavevector strongly resemble those modes in RMLs. Considering the similarity to RML coherent phonons, the overall decrease in transmission with increasing $N_s$ can be explained by our prior simulations on the length-dependence of coherent mode-conversion in RMLs \cite{maranets2025how}. Specifically, RMLs exhibit reduced transmission with increasing device length due to the finite spatial extensions of non-propagating coherent phonons that interact with the disorder of the device. In GMLs, increases in $N_s$ effectively raise the device length and contribute to attenuating transmission of these disorder-affected coherent phonons since they now have more disorder to interact with. We emphasize that this effect on coherent mode-conversion and phonon transmission suppression is separate from the behavior discussed in the previous section where increases in $N_p$ attenuates transmission due to the stimulation of periodic SL coherent phonons and not GML modes.

\subsection{Ascending vs. descending layer patterns\label{sec:asc_desc_pattern}}

\begin{figure}
    \centering
    \includegraphics[width=\textwidth]{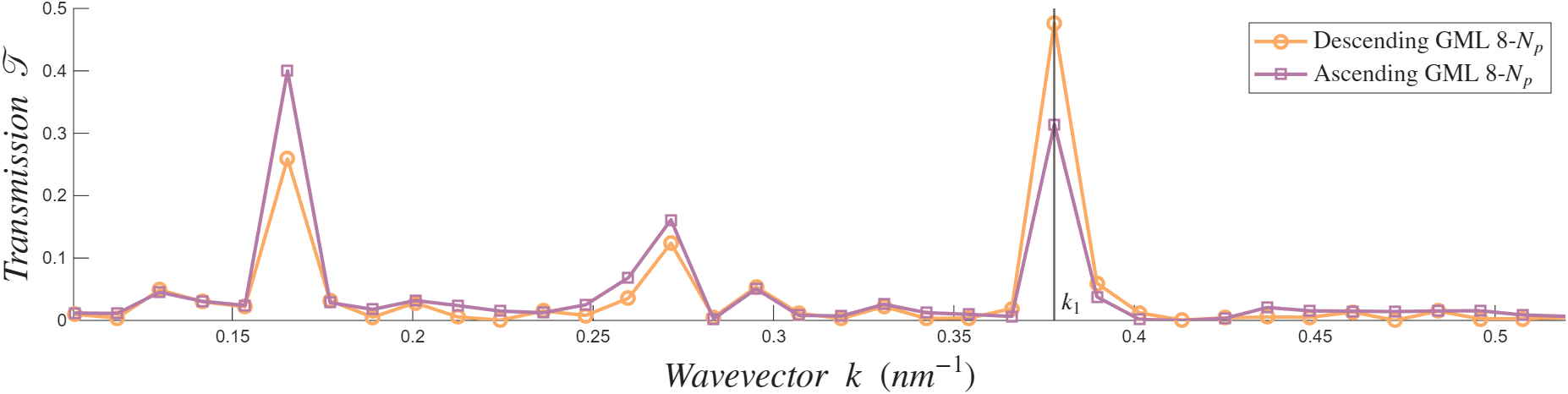}
    \caption{Transmission $\mathcal{T}$ versus wavevector $k$ for longitudinal-acoustic (LA) incoherent phonon wave-packets propagating through gradient multilayer device of $N_s=7$ and $N_p=8$ in ascending vs. descending layering patterns. A comparison of simulation domains is visualized in Fig.~\ref{fig:SL-GML-RML}a-\ref{fig:SL-GML-RML}b. Transmission is computed using Eq.~\ref{eqn:transmission_calc}.}
    \label{fig:ascendingvdescending}
\end{figure}

Besides the number of different period sizes $N_s$ and the number of periods for each period size $N_p$, we also examine if the layering arrangement of period sizes, in either an ascending or descending format, affects phonon transmission and coherent mode-conversion in GMLs. So far, we have only considered GML configurations of ascending layering patterns where the period sizes of the individual periodic SL domains incrementally increases along the axis of wave-packet propagation. In contrast, a descending layering pattern is a flipped configuration where the period sizes incrementally decrease. In Fig.~\ref{fig:ascendingvdescending}, we have plotted a comparison of the transmission spectra for an $N_s=7$ 8-$N_p$ GML with ascending and descending layering patterns. We find that the specific ordering of the interfaces has no significant effect on transmission, observing no changes in the spectra shapes and only slight changes in transmission magnitudes. Minimal difference between ascending vs. descending transmission through GMLs is further evidenced in wavelet transform calculations presented in Figs.~\ref{fig:gml_wavelet}c and ~\ref{fig:gml_wavelet}d. The coherent mode-conversion signatures of ascending and descending configurations are mirror images of each other, evidencing that the formation of coherent phonons in GMLs is determined by the distribution of period sizes (affected by both $N_s$ and $N_p$) and not the direction of period size gradient.

\section{Discussion\label{sec:Discussion}}

Unlike fully disordered RMLs, GMLs possess some level of order as determined by three structural parameters: (1) the number of different period sizes $N_s$, (2) the number of periods $N_p$ introduced for each distinct period size, and (3) the existence of either an ascending or descending layering pattern. Assessing how these parameters affect phonon transmission and coherent mode-conversion in GMLs using atomistic wave-packet simulations has unveiled two key insights about SL architectures and coherent phonon transport.

The first insight is an expected result that GMLs, being quasi-periodic, exhibit some behaviors that lie in an intermediate regime between behaviors observed in periodic and aperiodic SLs. In regards to the interface arrangement, GMLs possess short-range order but long-range disorder. When long-range disorder was low, effectuated by a low $N_s$, we observed coherent phonon behaviors comparable to those in periodic SLs. With increases in long-range disorder by raising $N_s$, the coherent phonons in GMLs transition to states more resembling the modes existing in RMLs. Clearly, coherent mode-conversion and thus transmission in GMLs straddles between the characteristics of periodic SLs and RMLs. Similar results was observed by Guo $et$ $al.$ \cite{guo2021thermal} who found GMLs exhibit partial phonon Anderson localization in contrast to RMLs which exhibit full localization and periodic SLs which exhibit no localization. Like our results, they also observed phonon transmission to decrease with increasing $N_s$, evidencing the influence of disorder on attenuating coherent transport.

The second insight is that long-range disorder influences phonon transmission and coherent mode-conversion far more than short-range order. In Sec.~\ref{sec:Number_of_periods} and Sec.~\ref{sec:asc_desc_pattern}, we investigated changes in $N_p$ and changes in the layering pattern (either ascending or descending, both of which affect short-range order, and neither manipulation had any considerable influence on transmission and coherent phonons. Though increases in $N_p$ effectuated reduction in transmission by effectively increasing the device length and long-range disorder, comparison to the spectra of RML devices of equivalent length showed minimal differences despite the fact that the GML possesses short-range order and the RML does not. Similarly, switching ascending vs. descending layering patterns in the GML corresponds to changes in in short-range order and this did not significantly impact transmission. It is possible short-range order has a more macroscopic influence considering GMLs and RMLs are observed to possess different lattice thermal conductivities \cite{chakraborty2020quenching}, however, our study reveals that short-range order is nearly irrelevant to the microscopic mechanisms of phonon coherence. These surprising findings underscore the importance of assessing phonon transport in SL devices through the framework of coherent mode-conversion. Analyses purely based on localization or scattering of periodic SL coherent phonons cannot rigorously explain effects such as the non-trivial thermal conductivities of RMLs and GMLs, the varied length-dependencies, and the negligible effect of short-range order on coherent phonon transmission uncovered in this study. The significance of coherent mode-conversion also emphasizes the immense value of atomistic wave-packet simulations as a most direct means to examine all facets of this phonon coherence effect.

\section{Conclusion\label{sec:conclusion}}

In summary, we have implemented atomistic phonon wave-packet simulations to study how the unique structural parameters of gradient SLs or GMLs possessing quasi-periodicity effectuate changes in phonon transmission and coherent mode-conversion from fully periodic SLs and fully aperiodic RMLs. We found that changes in long-range disorder, primarily influenced by manipulating the number of different period sizes $N_s$, profoundly impacts transmission and coherent mode-conversion. Specifically, the more disorder is introduced to the GML, the more coherent phonons begin to behave like the non-propagating modes of RMLs. However, short-range order, affected by changes in the number of periods added for each period size $N_p$ or the specific layering pattern  (either ascending or descending period sizes) is not observed to significantly impact transmission. These results suggest that long-range disorder is the dominant influence on spatial coherence behaviors in SL architectures.

\section*{Acknowledgements}
This work is supported by the National Science Foundation (Award Number: CBET-2047109). Doe is grateful for the partial financial support received from the Nevada NASA EPSCoR Undergraduate Research Opportunity Scholarship.

\section*{Author Contributions}
\textbf{Evan Doe:} Writing—original draft (lead), Writing—review \& editing (equal),  Software (lead), Methodology (lead), Formal analysis (equal), Conceptualization (equal). \textbf{Theodore Maranets:} Writing—review \& editing (equal), Writing—original draft (equal), Software (equal), Methodology (equal), Formal analysis (equal), Conceptualization (equal). \textbf{Yan Wang:} Writing—review \& editing (equal), Writing—original draft (equal), Supervision (lead), Funding acquisition (lead), Formal analysis (equal), Conceptualization (equal).

\section*{Data Availability}
The data that support the findings of this study are available from the corresponding
author upon reasonable request.

\section*{References}
\bibliographystyle{iopart-num} 
\bibliography{references}

\end{document}